# Piggybacking Codes for Network Coding: The High/Low SNR Regime

Samah A. M. Ghanem, *Senior Member, IEEE*

*Abstract*—We propose a piggybacking scheme for network coding where strong source inputs piggyback the weaker ones, a scheme necessary and sufficient to achieve the cut-set upper bound at high/low-snr regime, a new asymptotically optimal operational regime for the multihop Amplify and Forward (AF) networks.

## I. INTRODUCTION

Capacity of multicast linear networks is achievable by utilizing linear network coding [1]. The linear construction of network codes that achieves capacity [2], [3] in noisy wireless networks, or in noisy wired networks, might necessitate precoding and decoding designs that capitalize on connections between information flow measures and information reconstruction or estimation measures. Additionally, the use of different relaying protocols is at the heart of such goal.

In [4], Reznik et al. derive the optimal power distribution strategy among the transmitter and the relays that achieves capacity of a degraded Gaussian relay channel. In [5], Sankar et al. showed that Decode and Forward (DF) achieves the sum-capacity of degraded Gaussian Multiple Access (MAC) relay Channel. They showed that the MAC from source to relay is the bottleneck. One means to mitigate the MAC bottleneck is by the exploitation of Multiple Input Multiple Output (MIMO) techniques. In [6], Ekrem et al. proved the outer bound achievability for the capacity region of the degraded Gaussian MIMO broadcast channel utilizing tools that connects information to estimation measures [7].

In [8], Cover et al. showed that in a wireless network with single source-destination, compress-and-forward protocol achieves the cut-set bound, within a constant gap [9]. The claim was that this gap does not depend on channel gains, but it increases with the number of network nodes. In [10], Kramer showed that at high-snr regime, DF protocol exhibits a good scaling performance where the gap from the cut-set bound increases logarithmically with the number of nodes.

To have a linear network coding scheme that allows for closing the gap or mitigating the bottleneck in a wireless network with interference and noise, relays usually exploit the interference by forwarding it through the network to certain destination(s). Therefore, a natural and less complex strategy, is to amplify and forward the received sum of the noisy received signals, the so called analog network coding [11]. Gastpar et al. showed that uncoded transmission over two-hop amplify and forward can achieve the constant gap from the cut-set bound as the number of relays tends to infinity [12].

In [13], Maric et al. provided high-snr conditions under which multihop amplify and forward approaches capacity in a layered relay network. They showed that there exist a gap between the sum rate and the cut-set upper bound that is independent of channel gains.

In [14] and [15], Ghanem provided a generalized relationship that bridges connections between information flow measures or the mutual information (I) to estimation measures or the Minimum Mean Squared Error (MMSE), in a so called, "Multiuser I-MMSE", a relation that applies to multiuser Gaussian channels. In the same works, Ghanem provided a characterization of the derivative of the conditional and non-conditional parts of the mutual information. This included a characterization of the gap from the cut-set upper bound with respect to the channel, precoding and inputs estimates.

In principle, a user can be a source/sink terminal, and a multiple set of transmitting/receiving users correspond to multiple sources/sinks. Therefore, such relations open avenues[1] to address precoding strategies and operational regimes that are beyond ones limited to the high-snr asymptotically optimal regime for AF in multihop networks [13].

Therefore, using a similar framework of layered networks as in [13], and capitalizing on the multiuser/multiterminal I-MMSE [14], we provide an optimal transmit scheme adapted to the network level that provides a new asymptotically optimal operational regime of the AF, namely the high/low- mixed-snr regime.

The contributions of this paper are:

First, the proposal of a piggybacking scheme for the mutliterminal multihop AF network. This scheme is capacity achieving, energy efficient, bandwidth efficient, and provides relaxation on the synchrony between inputs. In particular, the scheme suggests, piggybacking low-snr inputs over high-snr ones, which can lead to having AF provide capacity for both. Therefore, the piggybacking scheme establishes piggybacking codes for network coding;

Second, we extend the optimality of AF protocol from being asymptotically optimal not only at the high-snr regime [13], but also optimal at a new interesting regime of high/low mixed-snr, getting around the necessity to use DF. Thus, the nodes will not necessarily decode then re-encode, but they can piggyback weaker inputs over strong ones, a strategy that achieves cut-set upper bound in the high-, and high/low-snr regimes;

Third, we shed light on the importance of the order of the estimation of inputs with different distribution than the Gaussian on the piggybacking scheme performance. In particular,

---

[1]The benefits of the "Multiuser I-MMSE" in [14] and [15] relation goes further beyond, to finding the capacity of interference channels, addressing the capacity of wireless networks, and to design interference-aware schemes, etc.

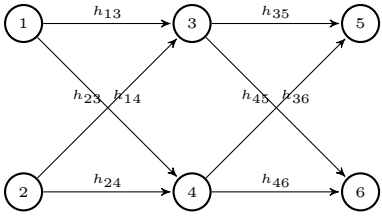

Figure 1. A two source two sink network. Sources 1 and 2 multicast independent data to nodes 5 and 6. Intermediate relays at nodes 3 and 4.

the piggybacking optimality is not affected by such order when inputs are Gaussian distributed;

Forth, the proposed piggybacking scheme provides an improvement on the snr of inputs facing low-snr conditions. Therefore, such strategy provides an energy efficient approach for networks where not all the power need to be used.

## II. Multihop Amplify and Forward

Consider the wireless network with two source two sink pair and two relays, with a deterministic network topology shown in Figure 1, that has a MAC channel output at node 5, given as follows,

$$y_5 = \sqrt{snr}h_{1,eq}x_1 + \sqrt{snr}h_{2,eq}x_2 + z_{eq} \quad (1)$$

where $x_j \sim \mathcal{N}(0, \mathbb{E}[x_j^2] = P_j), j = \{1,2\}$ is the channel input at node $j$, $h_{j,eq} = h_{35}h_{j3}\beta_3 + h_{45}h_{j4}\beta_4$ correspond to the channel gains of the network of the MAC at node 5, $z_{eq} = h_{35}\beta_3 z_3 + h_{45}\beta_4 z_4 + z_5$ is the noise due to amplified and forwarded noise plus the noise component at node 5 with $z_i \sim \mathcal{N}(0,1)$, $\beta_3, \beta_4$ are the amplification gains such that, $x_i(t) = \beta_i y_i(t-1)$, at intermediate nodes $i$ 3 and 4 respectively and $t$ is a timing index dropped from the rest of the paper. Similarly, we can write the channel output of the MAC at node 6.

To address AF schemes that can achieve the upper bounds of the MAC capacity at high- and high/low-snr regimes, we first state the achievable rates at node 5 as,

$$R_1 \leq I(x_1; y_5 | h_{1,eq}, h_{2,eq}) \quad (2)$$

$$R_2 \leq I(x_2; y_5 | x_1, h_{1,eq}, h_{2,eq}) \quad (3)$$

$$R_1 + R_2 \leq I(x_1, x_2; y_5 | h_{1,eq}, h_{2,eq}) \quad (4)$$

Then, we will introduce the tool that allows for understanding how such proposed scheme achieves the rates above with equality. In particular, to understand how the proposed piggybacking scheme works, we will utilize connections between information measures and estimation measures for the multiuser case. Such connections characterize changes on conditional and non-conditional mutual information rates as well as the joint rate of the MAC.

The optimization of key elements at the source nodes to maximize the joint mutual information will require joint estimation of the inputs. Therefore, of more practical relevance is to utilize a successive estimation process at the sink given the exploitation of conditional and non-conditional rate changes within those connections is feasible now [14]. The type of estimation of the inputs - linear/nonlinear - is driven by the inputs distribution.

First, we state the multiuser I-MMSE in [14], a fundamental relation between the derivative of the multiuser joint mutual information and the linear/non-linear MMSE with respect to the snr,

$$\frac{dI(snr)}{dsnr} = mmse(snr) + \psi(snr) \quad (5)$$

Where the total $mmse(snr)$ at node 5 is given by,

$$mmse(snr) = mmse_1(snr) + mmse_2(snr) \quad (6)$$

with per-input MMSE is given by,

$$mmse_j(snr) = \mathbb{E}\left[\|h_{j,eq}(x_j - \mathbb{E}[x_j|y_5])\|^2\right], \quad (7)$$

and the conditional mean estimator is defined as,

$$\mathbb{E}(x_j|y_5) = \sum_{x_j} \frac{x_j p_{y_5|x_j,x_k}(y_5|x_j,x_k) p_{x_j}(x_j) p_{x_k}(x_k)}{p_{y_5}(y_5)} \quad (8)$$

The conditioning on $x_k$ can be dropped if the estimation of $x_j$ is done considering the power of $x_k$ as noise[2].

We manipulate the multiuser/multiterminal I-MMSE to be suitable to the AF scheme where the noise variance rescales the snr evenly or unevenly according to the estimation of each input. Therefore, for the network in Figure 1 with AF we can rewrite (5) as,

$$\frac{dI(snr)}{dsnr} = mmse(\sigma_{z_{eq}}^{-1}snr) + \psi(\sigma_{z_{eq}}^{-1}snr) \quad (9)$$

where,

$$mmse(\sigma_{z_{eq}}^{-1}snr) = h_{1,eq}^2 P_1 E_1 + h_{2,eq}^2 P_2 E_2 \quad (10)$$

$$\psi(\sigma_{z_{eq}}^{-1}snr) =$$
$$- h_{1,eq}h_{2,eq}\sqrt{P_1}\sqrt{P_2}\mathbb{E}_{y_5}[\mathbb{E}_{x_1|y_5}[x_1|y_5]\mathbb{E}_{x_2|y_5}[x_2|y_5]^\dagger]$$
$$- h_{1,eq}h_{2,eq}\sqrt{P_1}\sqrt{P_2}\mathbb{E}_{y_5}[\mathbb{E}_{x_2|y_5}[x_2|y_5]\mathbb{E}_{x_1|y_5}[x_1|y_5]^\dagger]$$

and $\sigma_{z_{eq}}^{-1}$ is the inverse of the network noise variance which scales the snr of the input's estimates. The per-source network input Mean Squared Error (MSE) is given respectively as follows,

$$E_j = \mathbb{E}_{y_5}[(x_j - \mathbb{E}_{x_j|y_5}[x_j|y_5])(x_j - \mathbb{E}_{x_j|y_5}[x_j|y_5])^\dagger] \quad (11)$$

Therefore, taking the integral of both parts of (5),

$$I(snr) = \int mmse(\sigma_{z_{eq}}^{-1}snr)dsnr + \int \psi(\sigma_{z_{eq}}^{-1}snr)dsnr \quad (12)$$

The non-conditional and conditional components of the derivative of the mutual information, (see (14) and (15) in [14]) for the network of Figure 1 with AF, are given respectively as,

$$\frac{dI(x_1; y_5)}{dsnr} = mmse_1(\sigma_{eq}^{-1}snr) \quad (13)$$

and,

$$\frac{dI(x_2; y_5|x_1)}{dsnr} = mmse_2(\sigma_{z_{eq}}^{-1}snr) + \psi(\sigma_{z_{eq}}^{-1}snr) \quad (14)$$

Where $\sigma_{eq} = 1 + (h_{35}\beta_3)^2 + (h_{45}\beta_4)^2 + snr h_{2,eq}^2 P_2$ and $\sigma_{z_{eq}} = 1 + (h_{35}\beta_3)^2 + (h_{45}\beta_4)^2$.

---
[2]The conditioning on the channel is dropped, since the channel is considered deterministic and time invariant.

We define our optimization problems subject to per-source input power constraint as follows,

$$\max \quad I(x_1; y_5) \quad (15)$$

Subject to:

$$\mathbb{E}_x[x_1 x_1^\dagger] \leq P_1 \quad (16)$$

and,

$$\max \quad I(x_2; y_5 | x_1) \quad (17)$$

Subject to:

$$\mathbb{E}_x[x_2 x_2^\dagger] \leq P_2 \quad (18)$$

The optimization problems in (15) and (17) can be solved, applying the (Karush-Kuhn-Tucker) KKT conditions and capitalizing on the multiuser/multiterminal I-MMSE.

## A. Main Result: The high/low mixed-snr regime

We are interested in the regime in which one node transmits with high enough power so that the noise propagated by analog network coding is low. While the other input transmits with low enough power, so that it does not cause *destructive* but *constructive* interference[3], thus, the received snr is increased.

Definition: A wireless network is in the high/low- mixed-snr regime, if one node $k$ has, $\frac{1}{P_k} \geq \delta_k$, with $\delta_k \to 0$, and another node $j$ has, $\frac{1}{P_j} \leq \delta_j$, with $\delta_j \to \infty$. This implies that the received snr at the sink node $\ell$, has a high-snr $snr_\ell = (h_{j_{eq}}^2 P_j snr + h_{k_{eq}}^2 P_k snr)/\sigma_{z_{eq}}$.

Thus, such snr hits asymptotically the one for the MAC cut-set bound. At high/low-snr regime, since $P_1 \to \infty$, and $P_2 \to 0$, we avoid the bottleneck on the MAC that would be experienced if both inputs were at high-snr [13]. Therefore, if the first input with $\delta_1 \to 0$, at high-snr is scaled with $\sigma_{eq}$, while the second estimated input with $\delta_2 \to \infty$ at low-snr is scaled with $\sigma_{z_{eq}}$. This implies that, the transmit power of each input shall be different due to different scaling, and it is asymptotically expressed as a mixed regime of high/low-snr.

More clearly, to establish the operational asymptotic regime of "*high/low mixed-snr*" of the proposed piggybacking scheme: if one input at high-snr and another input at low-snr, the strong one could in effect piggyback off the other, thus getting around the necessity to decode and forward. It is instrumental to recall that such approach will allow for closing the gap between AF and the cut-set upper bound, i.e. achieves capacity as will be shown in the following section.

## III. Piggybacking Scheme

The proposed piggybacking scheme states that: if the strongest input piggybacks the other input, capacity can be achieved for both. The piggybacking strong input and the other input piggybacked are used as a code. Therefore, we refer to such codes as piggybacking codes for network coding.

More clearly, piggybacking is a pre-coding scheme that allows for joint access to the network equivalent MAC channel mitigating its bottleneck by creating constructively interfering signals in one signal with increased snr. Such increased snr signal convolves the strongest input with the other weaker input. This cooperative process is referred to as piggybacking where "The strong holds the weak" thus both are conveyed with no time-sharing.

Lets consider that both inputs to the multihop AF network $x_1$ and $x_2$ are Gaussian with zero mean and power constraints $E[x_1^2] = P_1$ and $E[x_2^2] = P_2$, respectively, contaminated along their flow in the network multihop AF by a Gaussian noise of variance $\sigma_{z_{eq}}$, as explained in the network model (1), such that at node 5,

$$R_1 + R_2 = \frac{1}{2} \log \left(1 + \frac{h_{1,eq}^2 P_1 snr + h_{2,eq}^2 P_2 snr}{\sigma_{z_{eq}}}\right) \quad (19)$$

The piggybacking scheme suggests that we can estimate $x_1$ while $x_2$ is treated as noise with respective rate,

$$R_1 = \frac{1}{2} \log \left(1 + \frac{h_{1,eq}^2 P_1 snr}{\sigma_{z_{eq}} + h_{2,eq}^2 P_2 snr}\right) \quad (20)$$

While the second input is estimated by assuming perfectly removing the knowledge of $x_1$, such that,

$$R_2 = \frac{1}{2} \log \left(1 + \frac{h_{2,eq}^2 P_2 snr}{\sigma_{z_{eq}}}\right) \quad (21)$$

The input (with high power) estimated first will piggyback the input (with low power) estimated next, thus allowing the sum of the rates, $R_1 + R_2$, to achieve capacity for both. Of particular relevance is to prove that the proposed piggybacking scheme captures the multiterminal I-MMSE behavior while yet optimal in achieving capacity. It is instrumental to know that for Gaussian inputs, the conditional mean estimators of inputs $x_1$ and $x_2$ given the output at node 5, are linear and given, respectively by,

$$\mathbb{E}_{x_1|y_5}[x_1|y_5] = \frac{\sqrt{snr_{high}}}{1 + snr_{high}} \tilde{y}_5 \quad (22)$$

Where $snr_{high} = \gamma \rho snr$, $\gamma$ is the snr scaling factor due to scaling input 1 with the variance of input 2 plus the noise variance, and $\rho = h_{1,eq}^2 P_1$. In turn, $\tilde{y}_5 = \sqrt{snr_{high}} x_1 + z$ a received signal scaled by input 2 variance plus the network noise variance, such that the noise $z$ is of unit variance. After complete removal of the estimated $x_1$, we have similarly,

$$\mathbb{E}_{x_2|y_5}[x_2|y_5] = \frac{\sqrt{snr_{low}}}{1 + snr_{low}} \hat{y}_5 \quad (23)$$

Where $snr_{low} = \zeta \nu snr$ and $\nu = h_{2,eq}^2 P_2$. In turn, $\hat{y}_5 = \sqrt{snr_{low}} x_2 + z'$ a received signal scaled by the network noise $\zeta$, such that the noise $z'$ is of unit variance. Therefore, the MMSE of input 1 and input 2, with the piggybacking scheme, are linear and given, respectively by,

$$E_j = \frac{1}{1 + snr_j} \quad (24)$$

In turn, substituting (22), (23), and (24) into (9),

$$\frac{dI(snr)}{dsnr} = \frac{h_{1,eq}^2 P_1}{1 + snr_{high}} + \frac{h_{2,eq}^2 P_2}{1 + snr_{low}} \quad (25)$$

---

[3]Constructive interference refers to the mutual interference introduced via cooperation to allow for canceling the effect of destructive non-mutual interference by increasing the SNR. An example on constructive interference introduced to interference channels is the MIMO channels or the cooperative interference channels.

Where, $\psi(\sigma_{z_{eq}}^{-1} snr) = 0$ due to orthogonality between input estimates, and due to complete removal of input 1 when estimating input 2. This implies that the change in the sum-rate associated to destructive non-mutual interference is mitigated at high/low-snr. In other words, the interference term in the multiterminal I-MMSE is canceled using the piggybacking scheme, which proves optimality of such scheme.

In general, it is worth to characterize such interference effect driven by the detection or estimation method. In particular, the interference effect or the rate loss (gap from the cut-set) due to the existence of input 2 as noise scaling the power of user 1, can be written as,

$$\int \psi(\sigma_{z_{eq}}^{-1} snr) dsnr = I(x_1; y_5) - I(x_1; y_5 | x_2)$$

$$= \frac{1}{2} \log\left(1 + \frac{h_{1,eq}^2 P_1 snr}{\sigma_{z_{eq}} + h_{2,eq}^2 P_2 snr}\right) - \frac{1}{2} \log\left(1 + \frac{h_{1,eq}^2 P_1 snr}{\sigma_{z_{eq}}}\right)$$

$$= \frac{1}{2} \log\left(1 + \frac{h_{1,eq}^2 P_1 snr + h_{2,eq}^2 P_2 snr}{\sigma_{z_{eq}}}\right)$$

$$- \frac{1}{2} \log\left(1 + \frac{h_{1,eq}^2 P_1 snr}{\sigma_{z_{eq}}}\right) - \frac{1}{2} \log\left(1 + \frac{h_{2,eq}^2 P_2 snr}{\sigma_{z_{eq}}}\right)$$

$$= I(x_1, x_2; y_5) - I(x_1; y_5 | x_2) - I(x_2; y_5 | x_1) \quad (26)$$

More clearly, within the context of the multiterminal I-MMSE, taking the derivative w.r.t the snr of both sides of the equation above, we have,

$$\psi(\sigma_{z_{eq}}^{-1} snr) = \frac{dI(x_1, x_2; y_5)}{dsnr} - \frac{dI(x_1; y_5 | x_2)}{dsnr} - \frac{dI(x_2; y_5 | x_1)}{dsnr}$$
$$= mmse_1(\sigma_{z_{eq}}^{-1} snr) + mmse_2(\sigma_{z_{eq}}^{-1} snr) - \psi(\sigma_{z_{eq}}^{-1} snr)$$
$$- mmse_1(\sigma_{z_{eq}}^{-1} snr) + \psi(\sigma_{z_{eq}}^{-1} snr)$$
$$- mmse_2(\sigma_{z_{eq}}^{-1} snr) + \psi(\sigma_{z_{eq}}^{-1} snr) \quad (27)$$

Therefore, if the difference between the sum-rates of the conditional mutual information and the joint mutual information is closed, we achieve the cut-set upper bound, or in other words, the derivative of the mutual information has the term $\psi(\sigma_{z_{eq}}^{-1} snr) \to 0$. In the following section, we will characterize the optimal power allocation of the piggybacking scheme that allows for closing the gap.

## IV. PIGGYBACKING SCHEME CHARACTERIZATION OF POWER ALLOCATION

The piggybacking scheme will lead to a mixed power allocation strategy. This is attributed to the high/low mixed-snr operational regime. In particular, the first estimated input should be assigned high power level to be able to scale it with the larger portion of the interfering signal plus noise variance, such that the estimation of the second input allows complete cancellation of the first input. In turn, a low power level assignment to the second input is sufficient to mitigate the left noise variance, and follows single user point-to-point channel assignment.

To characterize the piggybacking mixed power allocation strategy, we capitalize on the gradient of the non-conditional and conditional mutual information to find the optimal power allocation.

For Gaussian inputs, and according to the proposed piggybacking scheme, the gradient of the the non-conditional mutual information of the first estimated input with respect to $P_1$ is given by,

$$\nabla_{P_1} I(x_1; y_5) \sqrt{P_1} = \frac{1}{\sigma_{eq}} h_{1,eq}^2 P_1 E_1 snr = mmse_1(\sigma_{eq}^{-1} snr) \quad (28)$$

Therefore, the optimal power allocation for input 1 is given as,

$$P_1^* = \frac{\sigma_{eq}}{h_{1,eq}^2 snr} mmse_1^{-1}\left(\eta \frac{h_{2,eq}^2 P_2 snr + \sigma_{z_{eq}}}{h_{1,eq}^2 snr}\right), \eta < \frac{h_{1,eq}^2 snr}{\sigma_{eq}} \quad (29)$$

$$P_1^* = 0, \quad \eta \geq \frac{h_{1,eq}^2 snr}{\sigma_{eq}} \quad (30)$$

reduces to the waterfilling obtained by applying the KKT conditions that solves (15) and given as,

$$P_1^* = \frac{1}{\eta} - \frac{P_2 h_{2,eq}^2}{h_{1,eq}^2} - \frac{\sigma_{z_{eq}}}{h_{1,eq}^2 snr}, \eta < \frac{h_{1,eq}^2 snr}{\sigma_{eq}} \quad (31)$$

$$P_1^* = 0, \quad \eta \geq \frac{h_{1,eq}^2 snr}{\sigma_{eq}} \quad (32)$$

where $\eta^{-1}$ is the water level. It is clear that the power allocation has a term that accounts for scaling input 1 snr with the power of the second input, particularly, the second term of the right hand side of (31). On the other hand, the gradient of the conditional mutual information of the second estimated input with respect to $P_2$ is given by,

$$\nabla_{P_2} I(x_2; y_5 | x_1) \sqrt{P_2} = \frac{1}{\sigma_{z_{eq}}} h_{2,eq}^2 P_2 E_2 snr$$
$$- \frac{1}{\sigma_{z_{eq}}} h_{2,eq} h_{1,eq} P_1 \mathbb{E}_{y_5}[\mathbb{E}_{x_1 | y_5}[x_1 | y_5] \mathbb{E}_{x_2 | y_5}[x_2 | y_5]^\dagger] snr$$
$$= mmse_2(\sigma_{z_{eq}}^{-1} snr) \quad (33)$$

Given the orthogonality between linear estimates and the assumption of perfect reconstruction of the second input due to complete removal of first input, the second term in (33) which describes the gap, $\psi(\sigma_{z_{eq}}^{-1} snr) = 0$.

Therefore, the second piggybacked input optimal power allocation follows a single user mercury/waterfilling interpretation similar to the one in [16]. Following similar steps to the ones before, the optimal power allocation for input 2 is given as,

$$P_2^* = \frac{\sigma_{z_{eq}}}{h_{2,eq}^2 snr} mmse_2^{-1}\left(\eta \frac{\sigma_{z_{eq}}}{h_{2,eq}^2 snr}\right), \quad \eta < \frac{h_{2,eq}^2 snr}{\sigma_{z_{eq}}} \quad (34)$$

$$P_2^* = 0, \quad \eta \geq \frac{h_{2,eq}^2 snr}{\sigma_{z_{eq}}} \quad (35)$$

reduces to the single user waterfilling obtained by applying the KKT conditions that solves (17), and given as,

$$P_2^* = \frac{1}{\eta} - \frac{\sigma_{z_{eq}}}{h_{2,eq}^2 snr}, \quad \eta < \frac{h_{2,eq}^2 snr}{\sigma_{z_{eq}}} \quad (36)$$

$$P_2^* = 0, \quad \eta \geq \frac{h_{2,eq}^2 snr}{\sigma_{z_{eq}}} \quad (37)$$

It is straightforward to observe looking into (31) and (36) that the optimal power allocation of input 1 estimated first should exceed input 2 power, at such power set where one input piggybacks the other, we hit the cut-set upper bound. Moreover, its interesting to recall that we can yet hit the cut set upper bound if we increase the power of input 1 and 2 to be maximum. This resorts to the fact that the joint mutual information provides a waterfilling interpretation for input 2 similar to input 1 in (31) if input 1 is not removed when estimating input 2. This suggests that we can hit the cut-set upper bound with minimal energy via the piggybacking scheme, while yet keeping optimality when increasing the powers for the case of Gaussian inputs.

Worth to note that such piggybacking scheme defines a rate-splitting-like approach of two independent input's rates, where the interference of input 2 on input 1 is used as a code, and input 2 is a virtual input that is time-shifted in such a way that its signal convolves in time with input 1 data, which multiplies in the frequency band, allowing for bandwidth expansion as a coding gain, this is another way to understand the benefits introduced due to the piggybacking scheme [17].

## V. Optimality of Piggybacking scheme

In this section, we show the asymptotic optimality of the proposed piggybacking scheme at the high/low mixed-snr regime. To characterize the achievable rates with piggybacking at the high- and high/low-snr regimes, we recall the condition on the noise variance at high-snr given in [13] by,

$$\sigma_{z_{eq}} = 1 + \frac{h_{35}^2 P_3}{h_{13}^2 P_1 + h_{23}^2 P_2} + \frac{h_{45}^2 P_4}{h_{14}^2 P_1 + h_{24}^2 P_2} \quad (38)$$

We capitalize on (38) to study the high/low-snr, as a special case. It is clear that, when $P_1 \to \infty$ and $P_2 \to 0$, $\sigma_{z_{eq}} \to 1$. In turn, the first input achievable rate follows,

$$I(x_1; y_5) = \frac{1}{2} \log \left( 1 + \frac{h_{1,eq}^2 P_1 snr}{1 + h_{2,eq}^2 P_2 snr} \right) \quad (39)$$

Assuming perfect removal of linearly estimated $x_1$, the second input achievable rate follows,

$$I(x_2; y_5|x_1) = \frac{1}{2} \log \left( 1 + h_{2,eq}^2 P_2 snr \right) \quad (40)$$

According to the closed form of (39) and (40), the optimal power allocation yet follows (31)-(32) and (36)-(37). Therefore, we establish the operational regime of high/low mixed-snr at which piggybacking achieves the cut-set upper bound, with less energy consumption in the network.

Re-writing the joint mutual information of the MAC at node 5 by substituting the amplification factors as shown in (38) as $\beta_i = P_i/(h_{1,i}^2 P_1 + h_{2,i}^2 P_2), i = \{3, 4\}$ into the equivalent channel components, [13], we have,

$$I(x_1, x_2; y_5) = \frac{1}{2} \log \left( 1 + (h_{35}\sqrt{P_3 snr} + h_{45}\sqrt{P_4 snr})^2 \right) \quad (41)$$

Consequently, it is easy to observe that, at the high/low mixed-snr regime the sum rate at node 5 satisfies,

$$R_1 + R_2 > \frac{1}{2} \log \left( 1 + h_{35}^2 P_3 snr + h_{45}^2 P_4 snr \right) \quad (42)$$

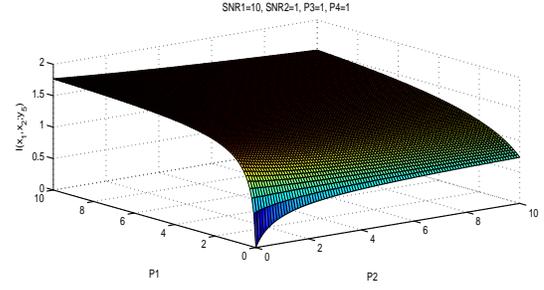

Figure 2. Joint Mutual Information with piggybacking: $I(x_1, x_2; y_5)$

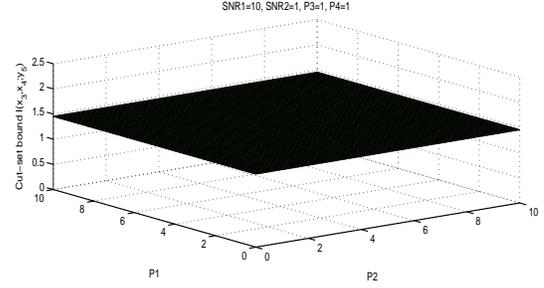

Figure 3. The cut-set upper bound at node 5: $I(x_3, x_4; y_5)$

However, the MAC cut-set bound at node 5,

$$I(x_3, x_4; y_5) = \frac{1}{2} \log \left( 1 + (h_{35}\sqrt{P_3 snr} + h_{45}\sqrt{P_4 snr})^2 \right) \quad (43)$$

Therefore, the cut-set upper bound is always achievable at the piggybacking operational asymptotic regime of high/low mixed-snr. If such conditions hold, the gap or the term $\psi(\sigma_{z_{eq}}^{-1} snr) = 0$ almost surely. If a degradation in the quality of the first input estimate occurred due to scaling its snr with a hugely amplified noise, this might lead to non-complete removal of the first input when estimating the second, thus, $0 < \psi(\sigma_{z_{eq}}^{-1} snr) \leq 0.5$ is bounded but not a constant gap as was known.

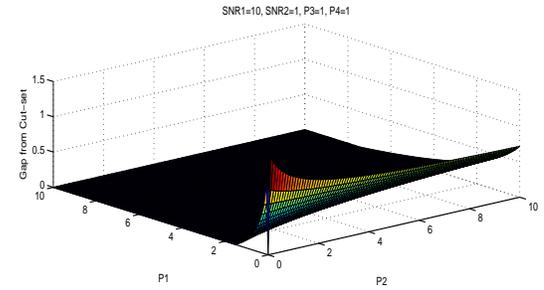

Figure 4. The gap from the cut-set bound at node 5

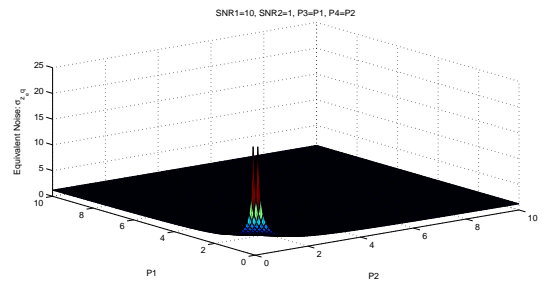

Figure 5. The network noise $\sigma_{z_{eq}}$ at node 5

Of particular relevance is to observe that the piggybacking necessitates a certain order of the inputs estimation at which the first estimated input must be the one with higher power, thus a low/high mixed-snr is not a candidate operational regime of the piggybacking scheme. However, for Gaussian inputs, it is straightforward to check that the order of estimation is not necessary due to the orthogonality of the input estimates.

## VI. SIMULATIONS

We shall now present a set of illustrative results that cast further insight into the proposed piggybacking scheme. In the following set of results, we use channel gains $h_{13} = 1$, $h_{14} = 1$, $h_{23} = 1$, $h_{24} = 1$, $h_{35} = 1$, $h_{36} = 1$, $h_{45} = 1$, $h_{46} = 1$, to isolate the effect of the channel gains from the impact of the study, particularly, we focus on the effect of the piggybacking of the inputs.

With high/low mixed-snr, we consider input 1 with high-snr $SNR1 = 10$ is estimated first and input 2 with low-snr $SNR2 = 1$ is estimated next. The amplification gains at intermediate nodes are dependent on source inputs powers and the powers at intermediate nodes 3 and 4, which are $P_3 = 1$ and $P_4 = 1$ respectively, a set of power levels at which we assure establishment of the operational regime without degrading the performance of the estimates.

We show in Figure 2 the joint mutual information at node 5 with successive estimation. Such successive estimation at node 5 is established for the piggybacking scheme with input 1, the strong input is piggybacking input 2, thus, input 1 is estimated-first, and the piggybacked input 2 is estimated-next accordingly. It is clear how the piggybacking scheme makes an efficient usage of the power in the system while achieving higher rates for both inputs, i.e. achieving capacity for both inputs. This can be clearly observed comparing the rate achieved at the power set $(P_1, P_2) = (10, 2)$, to the usage of maximum power at $(P_1, P_2) = (10, 10)$ which is associated with a decay in the achievable rate while yet hitting the cut-set bound.

Furthermore, the capacity achievability of the piggybacking scheme has been also demonstrated comparing Figure 2 with Figure 3-4, where the cut-set upper bound is achieved, and the gap is closed almost surely, respectively. The points in Figure 4 where the gap is above zero are those where the noise levels exceeds the input's snr in the scaling process of the first estimate, which is not part of the operational regime. Additionally, we can see from Figure 5, that the gap increases when the network noise variance is increased above unity, where the high/low-snr asymptotic condition $\sigma_{z_{eq}} \to 1$ is not met at $(P_1, P_2) = (0, 0)$, which are not part of the operational regime.

In turn, we can observe the necessity of the selection of which inputs to be piggybacking/piggybacked. This makes the order of estimation of the inputs at the sink fundamentally important to the performance of the piggybacking scheme. In particular, the order of the estimation defines a limiting factor on the noise scaling effect, thus the high-snr input is firstly estimated then the low-snr input. This is much relevant for arbitrarily distributed inputs rather than Gaussian distributed ones, where the estimation process is non-linear and the inputs estimates are non-orthogonal.

## VII. CONCLUSION

We show that piggybacking weak source inputs over strong ones will be necessary to achieve the cut-set upper bound of the multihop network with AF at high/low mixed-snr regime, a novel operational snr-regime at which AF provides optimality and achieves capacity. We shed light on the importance of the order of estimation to the optimality of the piggybacking scheme, albeit Gaussian inputs are insensitive to it.

## VIII. ACKNOWLEDGMENT

The author acknowledges fruitful discussions with Prof. Muriel Médard. I would also like to thank Prof. Médard for suggesting the piggybacking notion.